\documentclass[12pt]{iopart}

\usepackage{graphicx}
\usepackage{bm}        
\usepackage{amssymb}   

\begin{document}

\newcommand{\braket}[3]{\bra{#1}\;#2\;\ket{#3}}
\newcommand{\projop}[2]{ \ket{#1}\bra{#2}}
\newcommand{\ket}[1]{ |\;#1\;\rangle}
\newcommand{\bra}[1]{ \langle\;#1\;|}
\newcommand{\iprod}[2]{\bra{#1}\ket{#2}}
\newcommand{\logt}[1]{\log_2\left(#1\right)}
\def\cI{\mathcal{I}}
\def\cC{\tilde{C}}
\newcommand{\cx}[1]{\tilde{#1}}

 \title[Entanglement spectrum]{Entanglement spectrum and block eigenvalue spacing distribution of correlated electron states}
\author{Archak Purkayastha$^1$ and V. Subrahmanyam$^2$}
\address{International Centre for Theoretical Sciences, Tata Institute of Fundamental Research,  Bangalore-560012, India$^1$}
 \address{ Department of Physics, Indian Institute Of Technology,  Kanpur-208016, India$^2$}
 
 \ead{archakp@icts.res.in$^1$ and vmani@iitk.ac.in$^2$}
\date{\today}
\begin{abstract}
Entanglement spectrum of finite-size correlated electron systems are investigated using the Gutzwiller projection technique. The product of largest eigenvalue and rank of the block reduced density matrix, which is a measure of distance of the state from the maximally entangled state of the corresponding rank, is seen to characterise the insulator to metal crossover in the state. The fraction of distinct eigenvalues exhibits a `chaotic' behaviour  in the crossover region, and it shows a `integrable' behaviour at both insulating and metallic ends. The integrated entanglement spectrum obeys conformal field theory (CFT) prediction at the metal and insulator ends, but shows a noticeable deviation from CFT prediction in the crossover regime, thus it can also track a metal-insulator crossover. A modification of the CFT result for the entanglement spectrum for finite size is proposed which holds in the crossover regime also. The adjacent level spacing distribution of unfolded non-zero eigenvalues for intermediate values of Gutzwiller projection parameter $g$ is the same as that of an ensemble of random matrices obtained by replacing each block of reduced density matrix by a random real symmetric Toeplitz matrix. It is strongly peaked at zero, with an exponential tail proportional to $e^{-(n/R)s}$, where $s$ is the adjacent level spacing, $n$ is number of distinct eigenvalues and $R$ is the rank of the reduced density matrix.
\vskip 0.5cm
 \hfill {PACS: 03.67.Mn, 05.45.Mt, 71.30.+h, 71.27.+a}
\end{abstract}
\maketitle

\section{Introduction}

 Quantum entanglement of a system quantifies the correlations between the parts of the system\cite{Nielsen}, which serves as a resource for quantum information processing tasks.
The block entanglement, viz. the entropy of a subsystem,  is a widely-used  entanglement measure, that has been used to investigate critical behaviour near quantum phase transitions in spin systems\cite{plenio, sahoo}.  But there are not many studies of  the entanglement in interacting electron systems, which exhibit substantially richer structure than interacting spin systems as they carry additional charge degrees of freedom. In this article, we study the strong correlation effect on the entanglement spectrum of the 
one-dimensional Gutzwiller state\cite{Gutzwiller,metzner}, as a proto-type strongly-correlated state. 
\paragraph*{The Gutzwiller state :}
The Gutzwiller state was initially suggested as a variational ground state for the Hubbard model \cite{Gutzwiller}. In this state,  the strong on-site correlation effect of the Hubbard model ground state is mimicked by applying a  projection operator on the non-interacting metallic state to decrease the double occupancy. At one end of the control parameter is the metallic state with no projection, viz. a Fermi ground state constructed from occupying lowest-lying one-electron plane-wave states for both up and down spin electrons. The metallic state maximises the double occupancy as there is no correlation between the up and down spin electron. At the other extreme, there is an insulator phase, corresponding to the fully-projected state with no double occupancy. The Gutzwiller state for a lattice of $N$ sites is given by
\begin{equation}
| g\rangle = \prod_{i=1}^N\{1-(1-g)\hat{n}_{i\uparrow}\hat{n}_{i\downarrow}\}|F\rangle,
\end{equation}
where $|F\rangle =\prod_{k=0}^{k_F} \hat{c}_{k\uparrow}^\dagger \hat{c}_{k\downarrow}^\dagger |0\rangle$ is the metallic Fermi state constructed from the vacuum state by using electron creation operators $\hat{c}_{k\sigma}^\dagger$ with a momentum $k$ and spin $\sigma$, and $g$ is a parameter taking values from $0$ to $1$, $g=1$ being the non-interacting case, and $g=0$ being the limit of infinite interactions. The filling factor is determined by $k_F$,the Fermi momentum. 

While the Gutzwiller state gives good agreement with Hubbard model ground state for three dimensions, in 1D, it is different from the ground state of the Hubbard model. At half-filling, the ground state of the Hubbard model in 1D describes a Luttinger liquid, whereas the Gutzwiller state shows Fermi liquid behaviour \cite{metzner}. In the thermodynamic limit, for any $g \neq 0$, the momentum space distribution of electrons for the 1D half-filled Gutzwiller state has a discontinuity at the Fermi wave vector. Therefore, in the thermodynamic limit, the system is metallic for any $g \neq 0$. Also, in 1D, the half-filled Gutzwiller state for $g=0$ is the exact ground state of the Haldane-Shastry model \cite{Haldane Shastry}, which is a Heisenberg-like model with long range interactions.  Hence the Gutzwiller state is, by itself, an interesting correlated electron state. Here we will be interested in the entanglement spectrum of the half-filled Gutzwiller state. 

\paragraph*{Previous results : Entanglement entropy, fluctuations and metal-insulator crossover :\hfill}
The block entanglement entropy and bipartite fluctuations of the Gutzwiller state in 1D for half bipartition has been recently studied as a function of the correlation factor $g$ and the number of sites $N$ \cite{Archak}.The block entanglement entropy $S$ for half-bipartition of half-filled Gutzwiller state scales as :
\begin{equation}
\label{EE}
S = c_{eff}(g,N)(\frac{1}{2} + \frac{1}{3} log(N))
\end{equation} 
which has the same form as conformal field theory (CFT) result for one dimensional systems except with an effective central charge $c_{eff}(g,N)$ which is a function of both $g$ and $N$. At $g=1$, $c_{eff} = 2$ and at $g=0$, $c_{eff}=1$ which are independent of $N$ and are the correct results in these two limits from CFT. The $N$ dependence occurs for intermediate values of $g$. The scaling of $c_{eff}(g,N)$ indicates a metal-insulator crossover. Bipartite fluctuations also show a scaling. The scaling of both bipartite fluctuations and $c_{eff}(g,N)$ show that for $N<10^5$, the relevant scaling variable is $N^{1/3}g$ and metal-insulator crossover occurs at $N^{1/3}g\approx 0.24$.   

The reason behind deviation from the the CFT result for intermediate values of $g$ is the existence of a finite correlation length. As shown in Ref \cite{Archak}, the correlation length between opposite spins is infinte at $g=0$, but decreases with $g$ and becomes $0$ for $g=1$, whereas the correlation length between same spins always remain infinite. Hence, for intermediate values of $g$, there is still long range correlation in the system, but the system is no longer scale invariant because of the finite correlation length between opposite spins. So, CFT result for entanglement entropy with long range correlations is not valid, and also area law of entanglement entropy is not valid. We see that, in such case, the system still retains the $log(N)$ divergence of the CFT result for long range correlations, but with an effective central charge which is depends on $N$. 

For $g=0$ and $g=1$, the system is scale invariant and the CFT result for entanglement entropy with long range correlations holds with the appropriate central charge. 

It is important to note that the deviation from CFT result is only a finite size effect and will not be observed if the system is thermodynamically large. This is because the two relevant length scales in the system are the finite correlation length and the system size and the physics is governed by their ratio. In the thermodynamic limit, for any value of the finite correlation length, this ratio is $0$, and so, the system behaves like it is at $g=1$. Hence, in thermodynamic limit, the CFT result holds with central charge $c=2$ for  $g \neq 0$, and with $c=1$ for $g=0$. This is consistent with the known description for Gutzwiller state in thermodynamic limit, which says, it is metallic for all $g\neq 0$. 

For, finite size, however, the system shows a metal-insulator crossover when the ratio of the finite correlation length and system size is of order one, which occurs for a finite value of $g$. For $N<10^5$, this value of $g$ is given by $g\approx \frac{0.24}{N^{1/3}}$. Physical systems in this size range which may be described by the Gutzwiller state in one dimension are nanochains of strongly correlated metals.

\paragraph*{Entanglement spectrum:}
All the above previous results are based on the study of entanglement entropy and bipartite fluctuations. While entanglement entropy is a number characterising the correlations and entanglement in the system, the entanglement spectrum, which is the full spectrum of eigenvalues of the reduced density matrix, is more fundamental. More information regarding correlations and entanglement is expected to be there in the entanglement spectrum. Also, in numerical techniques based on matrix product states, like the density matrix renormalization group, all observables are calculated from a truncated spectrum of the reduced density matrix. Thus study of full spectrum of the reduced density matrix also helps in determining convergence and accuracy for such numerical schemes. In this paper, we investigate the spectral features of the reduced density matrix of the finite size 1D half-filled Gutzwiller state for half bipartition. Entanglement spectrum has been previously studied for 1D systems within the framework of CFT \cite{CFT1}. It is interesting to see how these results are modified for our case, where we know CFT is not valid.

The results are based on exact numerical Schmidt decomposition of systems upto $N=16$ sites. We consider a  ring with $N$ electrons, with equal number of up and down spins, corresponding to the half-filling case. We partition the system into two equal parts, with the first $N/2$ sites belong to one subsystem. In the following, we keep the system size at $N=16$ and study various observables as a function of $g$. Finite-size scaling is difficult to be implemented as the data is insufficient. However, since, as described above, the correct scaling variable is a product of an increasing function of $g$ and an increasing function of $N$, we expect that the variation of observables with $g$ for fixed $N$ is representative of variation of observables with $N$ for a fixed $g$.

\section{Largest eigenvalue and rank}

\begin{figure}[t]
\includegraphics[scale=0.45]{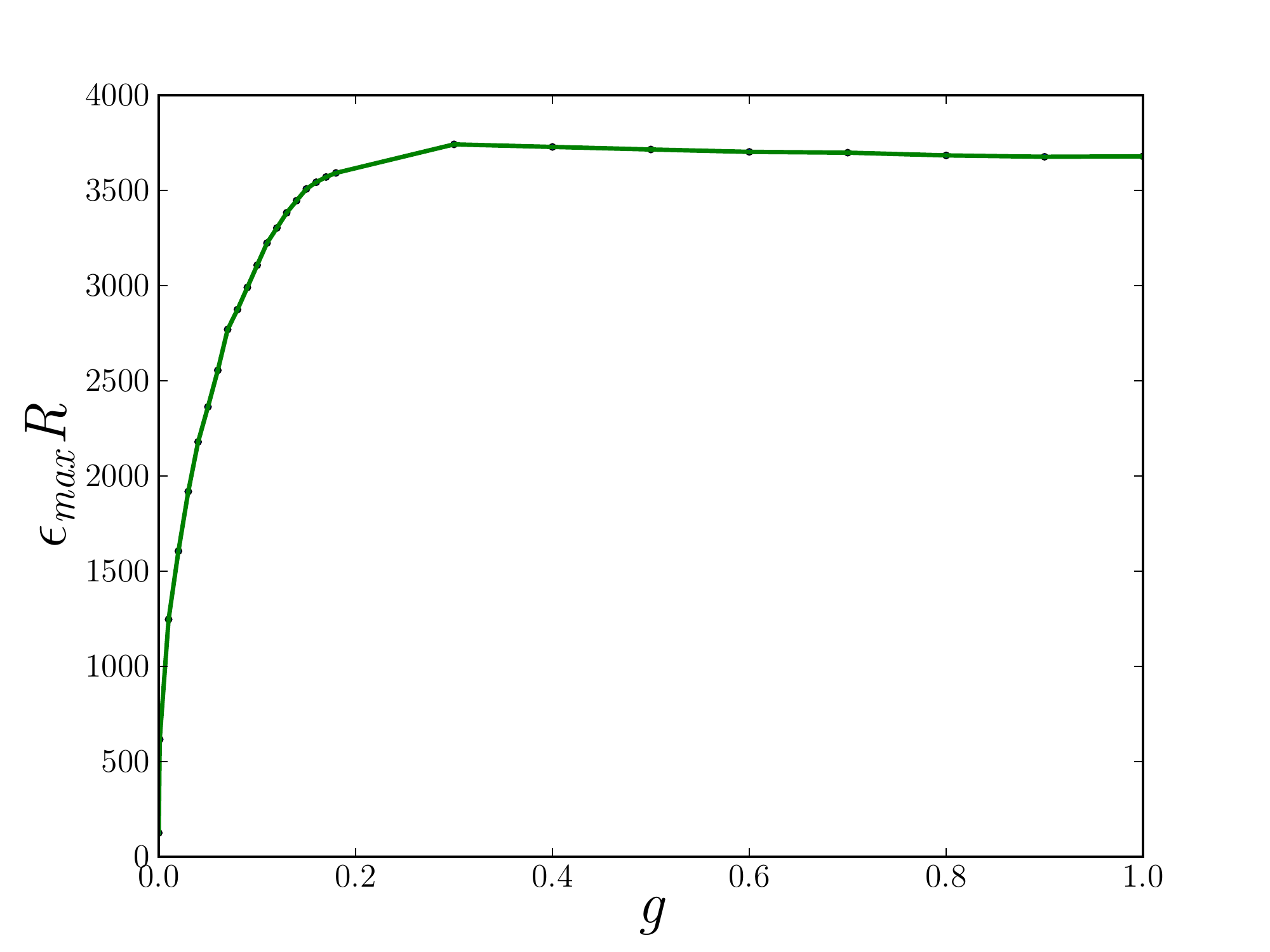}\llap{\makebox[7cm][t]{\raisebox{1.2cm}{\includegraphics[width=6cm,height=4cm]{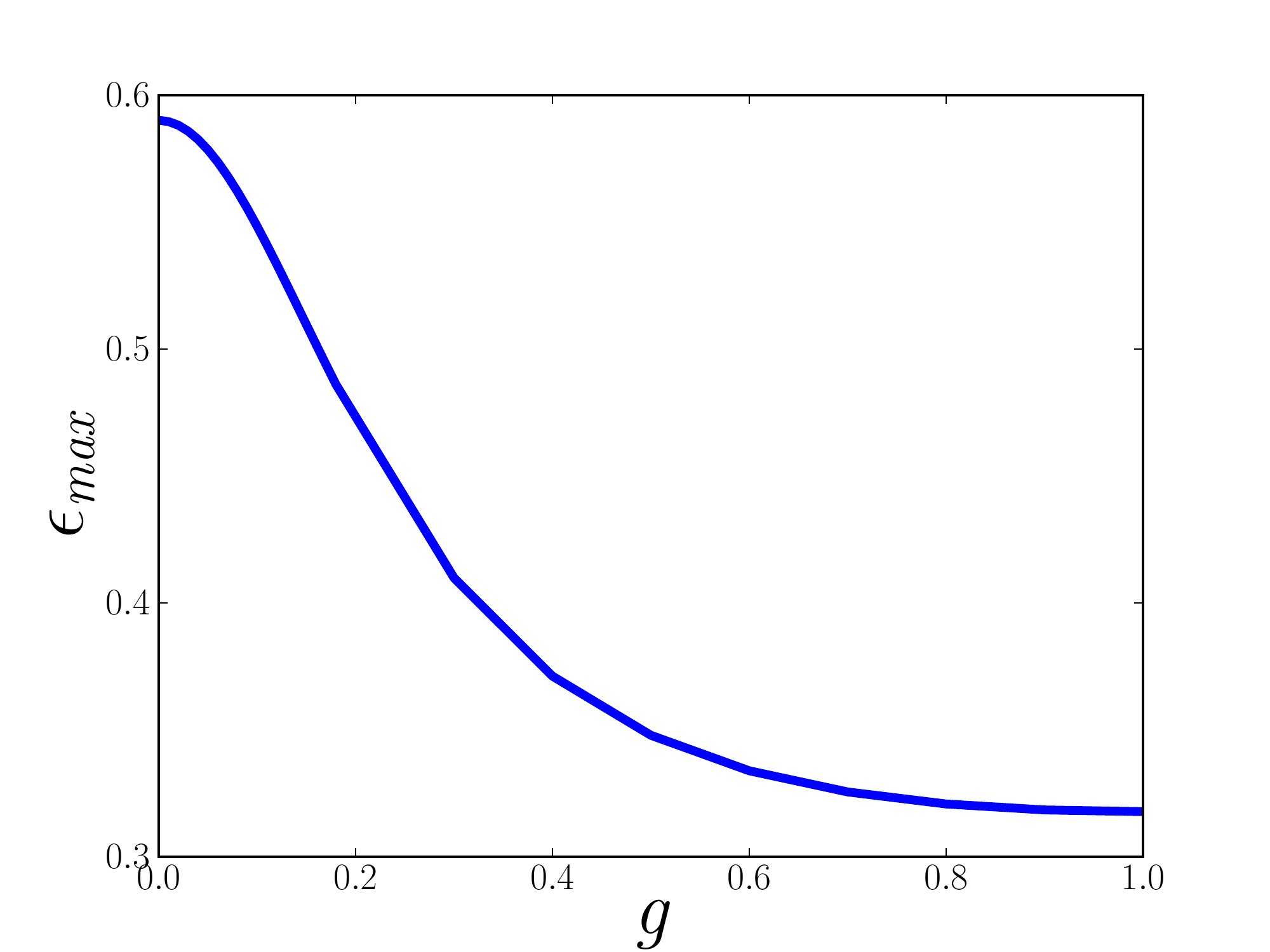}}}}
\caption{\label{fig:emaxR} The variation of the product of the largest eigenvalue $\epsilon_{max}$ and rank $R$ of the reduced density matrix with on-site correlation factor $g$ for $N=16$ sites at half-filling is shown.  The inset shows the variation of $\epsilon_{max}$ with $g$. Crossover to metallic limit is shown by the product becoming almost constant.}
\end{figure}

The dimension of the reduced density matrix of the subsystem with $N/2$ sites  is $2^{N}$, and in principle as many eigenvalues can be nonzero.  But, the entanglement entropy shows only a logarithmic divergence with $N$, implying only O($N$) number of nonzero eigenvalues. Thus the rank of the reduced density matrix is vastly reduced. Since on-site correlation factor $g$ reduces the probability of doubly-occupied sites,  the rank is reduced  as $g$ decreases. Reduction of the rank  of the reduced density matrix is one of the reasons for a decrease of the entanglement as $g\rightarrow0$.

The largest eigenvalue is an important parameter governing the distribution of eigenvalues. The range of the distribution of eigenvalues $\{\epsilon_i\}$ is determined by the largest eigenvalue $\epsilon_{max}$. For Gutzwiller state at half-filling, $\epsilon_{max}$ decreases as correlation factor $g$ increases. This is because, as the rank of the reduced density matrix increases, the normalisation factor demands a decrease of $\epsilon_{max}$. 

For a given rank $R$, the minimum value of $\epsilon_{max}$ is $1/R$, which corresponds to the maximally entangled state where all eigenvalues are equal. Thus minimum value of the product $\epsilon_{max}R$ is $1$. Hence as the product increases from $1$, it gives a measure of how far the system is from the maximally entangled state of the corresponding rank.

In  Fig.\ref{fig:emaxR},  we see that the product $\epsilon_{max} R$ is $>>1$ and increases rapidly from  the insulating limit and saturates at the metallic limit at $g=1$. This means that the state is always far from the maximally entangled state. But the strongly interacting case is closer to the maximally entangled state of the corresponding rank, than the metallic state, even though the metallic state has greater entanglement entropy (see introduction, Eq \ref{EE}) due to its greater rank. The onset of the crossover to metallic region is seen by the product becoming almost constant. The crossover region shown by this curve matches with that from the scaling form for the entanglement entropy\cite{Archak}.  Hence distance from the maximally entangled state also characterises the crossover. 

\begin{figure}[t]
\includegraphics[scale=0.45]{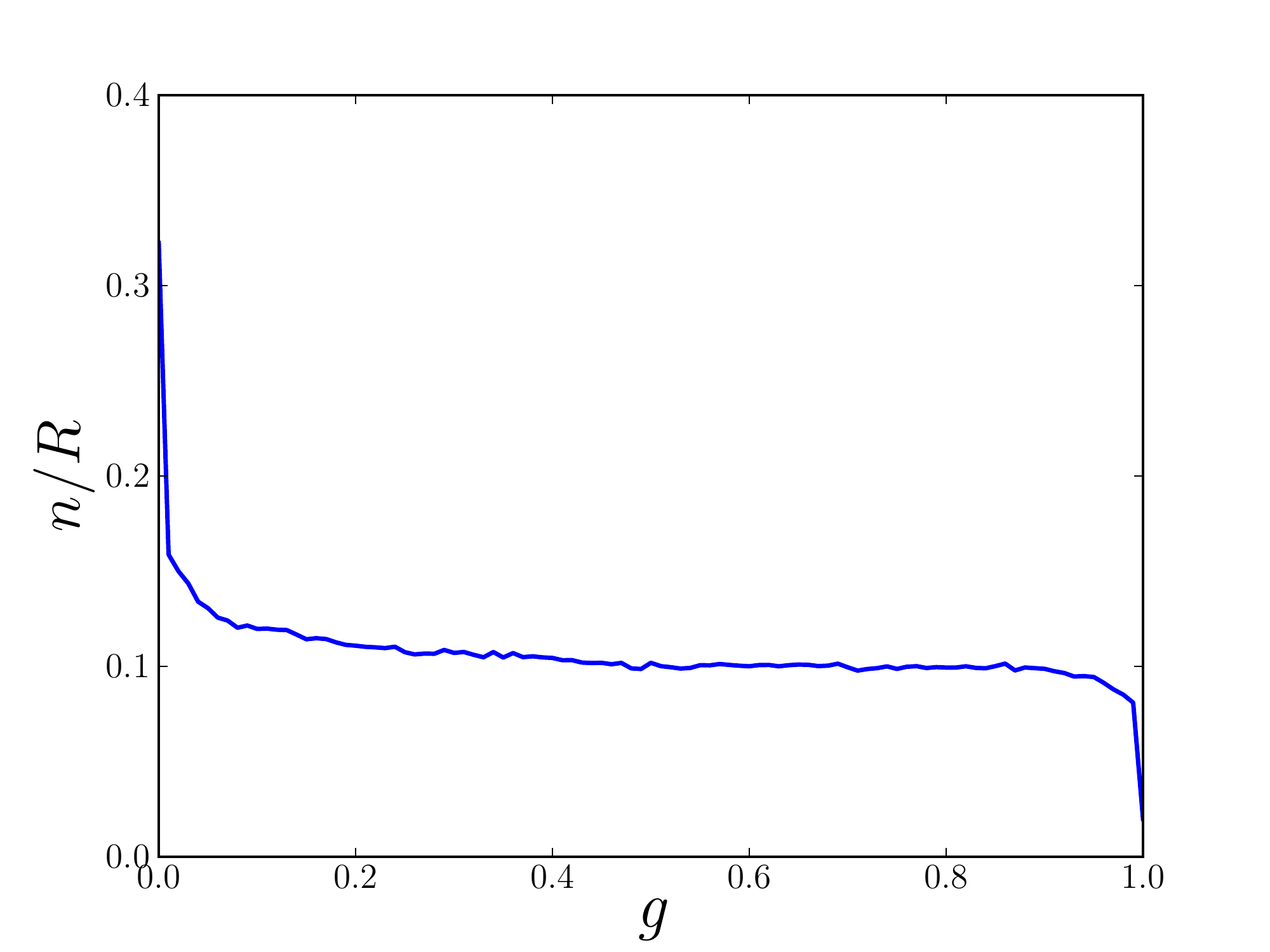}
\caption{\label{fig:n/R} Variation of ratio of number of distinct eigenvalues $n$ to rank $R$ with $g$ for $N=16$ sites at half-filling is shown. It remains nearly constant, exhibiting a `chaotic' behaviour in the crossover region, but it decreases with $g$ close to $g=0$ and $g=1$, showing an `integrable' behavior.}
\end{figure}

\begin{figure*}
\includegraphics[width = 16cm, height = 11cm]{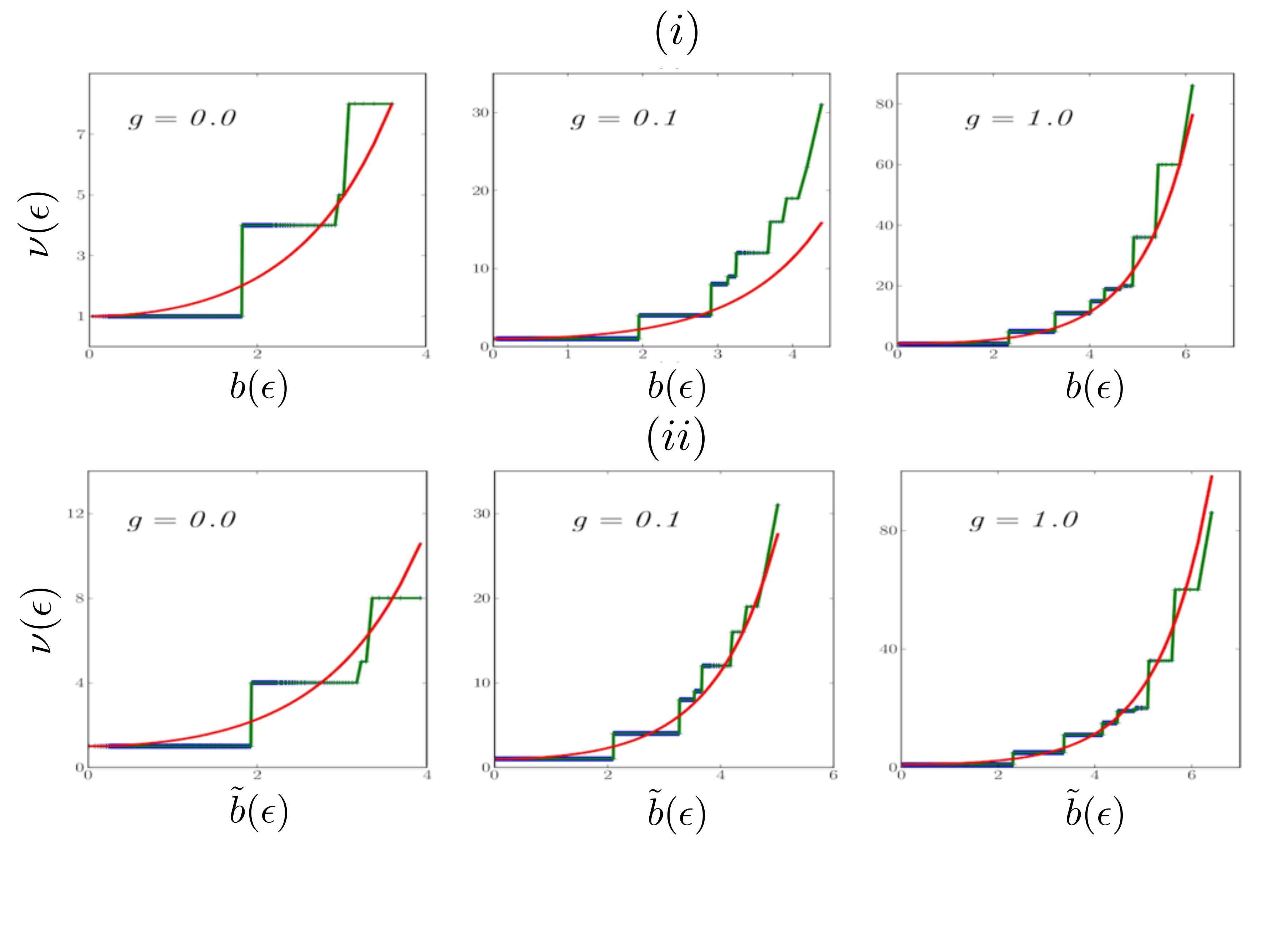}
\caption{\label{fig:n(e)1} The staircase plots show the integrated density of eigenvalues $\nu(\epsilon)$ for $N=16$ sites at half filling for various $g$ values $(i)$ with $b(\epsilon)=2\sqrt{ln(\epsilon_{max})ln(\frac{\epsilon}{\epsilon_{max}}})$, along with the corresponding CFT (smooth curve) result given in Eq.\ref{nu},$(ii)$ with $\tilde b(\epsilon)=2\sqrt{\mid \frac{S}{2}(ln \hspace{2pt} \epsilon+\frac{S}{2})\mid}$, along with the modified  CFT (smooth curve) result given in Eq.\ref{nu2}. $g=0.0$ corresponds to the insulating state, $g=0.1$ ($\simeq 0.24/N^{\frac{1}{3}}$ ) corresponds to the crossover, $g=1.0$ corresponds to the metallic state. 
}
\end{figure*}

\section{Fraction of distinct eigenvalues}

Fraction of distinct eigenvalues can be used to track quantum chaos. In study of quantum chaos, a Hamiltonian describing a chaotic system is one which can be modelled by a random matrix ensemble with the same degree of freedom as the Hamiltonian, viz. both have the same spacing distribution given by the Wigner surmise (the GOE, GUE or GSE spacing distribution) \cite{Bohigas}. The Wigner surmise gives a zero at zero spacing, which means there can be no degeneracy of eigenvalues. For an integrable system, however, the spacing distribution is given by a Poisson distribution, which has a peak at zero spacing, which means a large number of eigenvalues are degenerate. If we have an integrable system and then introduce a symmetry breaking perturbation via a control parameter that makes it non-integrable, it has been recently shown \cite{Ranjan} that the spacing distribution changes from Poisson to the Wigner surmise. Therefore, the fraction of distinct eigenvalues changes from less than one to one as the symmetry-breaking parameter is increased and then remains constant at one. Thus, the fraction of distinct eigenvalues remaining constant as the symmetry-breaking parameter is changed can be taken as a signature of quantum chaos.    

Instead of the Hamiltonian, here we are concerned with the reduced density matrix of bipartition of the Gutzwiller state, but, we have a similar situation. For $g=0$ and $g=1$, because of scale invariance, the system is amenable to analytical methods via CFT, and thus is `integrable', whereas for intermediate values of $g$, scale invariance is broken and it becomes `non-integrable'. The parameter $g$ therefore acts like the symmetry-breaking parameter.

Fig \ref{fig:n/R} shows the variation of the fraction of distinct eigenvalues with $g$ for  the reduced density matrix of half-filled Gutzwiller state. We see that for intermediate values of $g$, the fraction of distinct eigenvalues remain almost constant. This then may be a signature of quantum chaos. However, this constant is much smaller than one. This is because, even though scale invariance is broken, the symmetries and the block diagonal nature of the reduced density matrix preserves a lot of degeneracy of the eigenvalues. That this is indeed a signature of quantum chaos will be made clear from the detailed analysis of the spacing distribution (Section \ref{Spacing distribution}.). The detailed description of the various symmetries of the reduced density matrix will also be given while studying the spacing distribution.

Near $g=1$, the fraction of distinct eigenvalues decreases from the constant value because  degeneracy increases, while the rank remains nearly the same. On the other hand, near $g=0$, the fraction of distinct eigenvalues increases from the constant value because of a drastic decrease in rank. 


\section{\label{ESGS}Integrated entanglement spectrum }

Next we look at the density distribution of non-zero eigenvalues. The block entanglement entropy can be written, using the density of eigenvalues, as
 
\begin{equation}
S  = -\sum_i \epsilon_i \hspace*{3pt} ln \hspace*{2pt} \epsilon_i  
   = \int_0^1 d\epsilon \hspace{4pt}(1+ \ln  \epsilon )\hspace{4pt}(R- \nu (\epsilon)) 
\end{equation}
Here, the integrated density of eigenvalues $\nu(\epsilon)$ is given by
\begin{equation}
\label{theta}
\nu(\epsilon) = \sum_i\Theta (\epsilon_i - \epsilon).
\end{equation}

Calabrese and Lefevre\cite{CFT1} calculated this function  for one-dimensional gapless systems using CFT. They started from the result obtained by scaling analysis for large subsystem size:
\begin{equation}
\label{Ralpha}
Tr \rho ^ {\alpha} = c_{\alpha}e^{-b_0(\alpha - 1/\alpha)}, 
\hspace{20pt} b_0  = \frac{c}{6} log(L) 
\end{equation}
where $c$ is the central charge of the CFT, and $L$ is the subsystem size. Taking the coefficient $c_{\alpha}$ to be constant, they calcualted the density distribution of eigenvlaues of the reduced denstity matrix and from there obtained the result for the integrated density of eigenvalues as :
\begin{equation}
\nu(\epsilon) = I_0(2\sqrt{-b_0ln(\frac{\epsilon}{\epsilon_{max}}}) ) \label{nu}
\end{equation}
where $I_0$ is modified Bessel function of zeroth order, $\epsilon_{max}$ is the largest eigenvalue. Various authors have shown \cite{CFT3} :
\begin{equation}
\label{b01}
b_0  \simeq -ln \hspace{2pt} \epsilon_{max} 
\end{equation}
in the large $L$ limit. Also, following Cardy and Calabrese \cite{CFT2}, entanglement entropy is given by :
\begin{equation}
S = - \lim_{\alpha \rightarrow 1} \frac{\delta}{\delta \alpha} Tr \rho ^ {\alpha} = 2b_0 = \frac{c}{3} log(L)
\end{equation}
which is also true in the large $L$ limit. Above statement is just restatement of the result that entanglement entropy equals the single-copy entanglement in the large $L$ limit.

In  Fig.\ref{fig:n(e)1}($i$),  the integrated eigenvalue density $\nu$ is plotted as a function of $b=2\sqrt{ln(\epsilon_{max})ln(\frac{\epsilon}{\epsilon_{max}}})$ for $N=16$ sites for various values of on-site correlation factor $g$, along with the corresponding plots of the CFT result given above. The CFT result is seen to be correct only when $g=0$ and $g=1$ and not for intermediate values of $g$. This, of course, as described in introduction,  is not surprising in light of previous work \cite{Archak}. There it is shown that CFT results are valid only at the two end points. The reason is that it is only at the two end points that all underlying correlation lengths are infinite. For intermediate values of $g$, there is a finite correlation length that comes from correlation between opposite spins. Hence the system does not remain scale invariant and finite-size corrections become important. Deviations from CFT result signifies the crossover. However, the entanglement entropy retains CFT-like scaling only with a different effective central charge which includes finite-size corrections. So, we expect that upon replacing the central charge in above equations by the effective central charge will give the correct $\nu$ for all values of $g$. Equivalently, we can just go back to the result $b_0 = \frac{S}{2}$. So, the formula for integrated eigenvalue spectrum becomes :
\begin{equation}
\nu(\epsilon) \simeq I_0(2\sqrt{ \frac{S}{2} \mid ln \hspace{2pt} \epsilon+ \frac{S}{2} \mid} \hspace{2pt} )  \label{nu2}
\end{equation}
where the modulus is inserted because the term inside is no longer guaranteed to be positive. We have shown in Fig.\ref{fig:n(e)1}($ii$) the entanglement spectrum staircase plotted with the modified function given above. It is seen that Eq.\ref{nu2} holds for all values of $g$. This confirms Eq. \ref{Ralpha} holds for intermidiate $g$ also but with an effective finite-size corrected central charge.   

\begin{figure*}
\includegraphics[width=19cm, height=5.7cm]{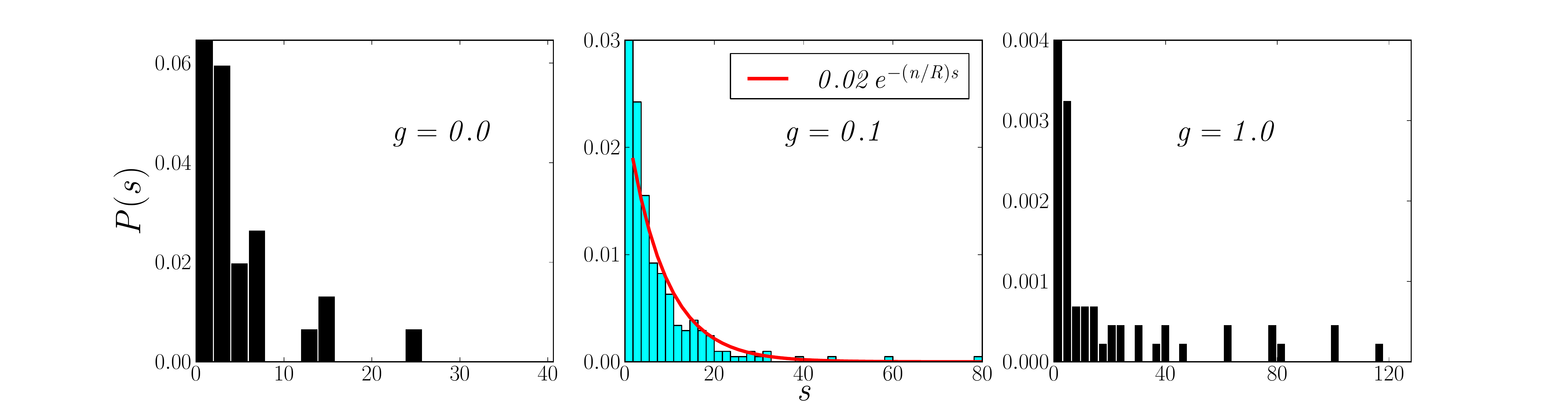}
\caption{\label{fig:P(s)1} Spacing distribution of the unfolded eigenvalues of $\rho_A$ is plotted with the adjacent level spacing $s$, for $N=16$ sites at half-filling for various values of $g$. The height of the big peak at $s=0$ for all $g$ is not shown in the figure as it would wash out the structure at other values of $s$. At intermediate values of $g$, where is system is `chaotic', the tail of the distribution decays exponentially as $e^{-(n/R)s}$ where $n$ is the number of distinct eigenvalues and $R$ is the rank of $\rho_A$. This is as expected from random matrix theory.  The spacing distribution at the two ends $g=0$ and $g=1$ has a different nature because the system is `integrable' at those values of $g$. }
\end{figure*}

\section{Spacing distribution} \label{Spacing distribution}

Spacing distribution of eigenvalues of Hamiltonian systems has been traditionally used to investigate the quantum chaos\cite{Bohigas}, and to track metal-insulator transition\cite{spacing1}. The uncorrelated metallic state shows a `correlated' single-particle energy level spacing distribution with level repulsion, whereas the correlated insulator state is associated with a `uncorrelated' Poisson spacing distribution. For a chaotic system, the spacing distribution is seen to be same as that of an ensemble of random matrices having the same number of degrees of freedom as the system. From our investigation of fraction of distinct eigenvalues we have found that for crossover values of $g$, the eigenvalues of reduced density matrix behave like a `chaotic' system. So, in this region, we expect that the spacing distribution of eigenvalues of $\rho_A$ can be understood from a random matrix ensemble.

For that purpose, we first look at the symmetries of $\rho_A$. $\rho_A$ is block diagonal, with each block specified by the number of up spins $N_{A\uparrow}$ and the number of down spins $N_{A\downarrow}$ in $A$. Furthermore, we note that the labels of $\uparrow$ and $\downarrow$, as well as those of $A$ and $B$ are arbitrary and can be interchanged. Hence the eigenvalues of blocks of reduced density matrix with spins flipped are same. Also, all the eigenvalues from blocks with $N_A$ particles in $A$ are same as those from blocks with $N-N_A$ particles in $A$. Hence there are inherent degeneracies among the blocks.

Finding eigenvalues of each block of $\rho_A$ is a formidable task because the size and number of blocks increase exponentially with $N$. Let  $R$ denote a block of the reduced density matrix $\rho_A$ with $N_{A\uparrow} = p$ and $N_{A\downarrow} = q$. The  matrix element  $R_{ij}$ between two configurations labelled by the set of positions of up and down spin electrons in subsystem $A$ is given by,

\begin{equation}
\label{R1}
R_{ij} =  Q \hspace*{2pt} f^A_{ij}(g) 
 \sum_{b_{\uparrow},b_{\downarrow}} f^B_{b_{\uparrow} b_{\downarrow}}(g) \times 
D_{k}(i_{a\uparrow},b_{\uparrow})D_{k}(i_{a\downarrow},b_{\downarrow}) 
D^*_{k}(j_{a\uparrow},b_{\uparrow})D^*_{k}(j_{a\downarrow},b_{\downarrow}),
\end{equation}  

where $Q$ is constant that depends on $N$, $N_{A\uparrow}$ and $N_{A\downarrow}$, $f^A_{ij}(g)$ is a function of $g$ coming from the double occupancy of the configurations $i$ and $j$  of subsystem $A$, $f^B_{b_{\uparrow} b_{\downarrow}}(g)$ is a function $g$ coming from the double occupancy of the configuration of subsystem $B$. The  label $i_{a\uparrow}$ ($i_{a\downarrow}$) is a set of $p$ ($q$) numbers giving positions of up (down) spins in subsystem $A$ in $i$th configuration  and similarly for the $j$th configuration, and $b_{\uparrow}$ ($b_{\downarrow}$) is the set of positions of up (down) spin in subsystem $B$, and $D_{k}(r)$ is the Slater determinant with momentum labels given by the set $k$ and position labels given by the set $r$, $*$ denotes complex conjugate. The sum is over all possible configurations of $B$, i.e, all possible choices of $b_{\uparrow}$ and $b_{\downarrow}$. Writing out the Slater determinants explicitly and simplifying, it can be shown that there are some terms in the sum that are independent of the configurations of $B$. Upon performing the sum over configurations of $B$, these terms add up coherently whereas the other terms add incoherently. Since the number of all possible configurations of $B$ is quite large, we can approximate $R_{ij}$ by considering only the terms independent of configurations of $B$. These terms will only pick up a factor of some constant times a function of $g$ from the sum.  The result is of the form :
\begin{equation}
R_{ij} \simeq Q \hspace*{2pt} f_{ij}(g) \hspace*{2pt}  \hspace*{2pt}
[\sum_{k_p} D_{k_p}(i_{a\uparrow})D^*_{k_p}(j_{a\uparrow}) 
\sum_{k_q} D_{k_q}(i_{a\downarrow})D^*_{k_q}(j_{a\downarrow})]
\equiv Q f_{ij}(g) M_{ij},  
\end{equation} 
where all constants have been absorbed into $Q$, all $g$ dependence coming from both subsystems $A$ and $B$ has been absorbed into $f_{ij}(g)$, $k_p$ ($k_q$) is a set of $p$ ($q$) values of $k$ chosen from all possible values of $k$ below $k_F$ and the sums are over all possible such choices, $M_{ij}$ is the term within square brackets which is the sum of all possible terms in Eq \ref{R1} which are independent of configurations of $B$.  Since taking complex conjugate of Slater determinant simply means $k \rightarrow -k$, and since both these levels are allowed, the above sums are real. Thus the $M_{ij}$ is real. Also due to the same reason, $M_{ij}$  depends only on the relative positions of up (down) spins in the $i$th and $j$th configurations. These properties manifest in symmetries of the matrix $M$. For $N=4$, explicit calculation shows that the matrix $M$ is a real symmetric Toeplitz matrix.For example, for $N=4$, for the block with $N_{A\uparrow} = 1$ and $N_{A\downarrow} = 1$, $M_{ij}$ is given by :
\begin{equation}
M_{ij} =  1+cos\frac{\pi}{2}(i_{a\uparrow}-j_{a\uparrow})+cos\frac{\pi}{2}(i_{a\downarrow}-j_{a\downarrow}) 
 +cos\frac{\pi}{2}(i_{a\uparrow}-j_{a\uparrow}+i_{a\downarrow}-j_{a\downarrow})
\end{equation}
which can be cast into Toeplitz form with proper ordering. For $N=6$, the matrix $M$ becomes block real symmetric Toeplitz matrix with each block also being a real symmetric Toeplitz matrix. Such a matrix has the same number of degrees of freedom as a real symmetric Toeplitz matrix. However, it is difficult to prove such a result for general $N$ directly. 

\begin{figure}
\includegraphics[scale=0.45]{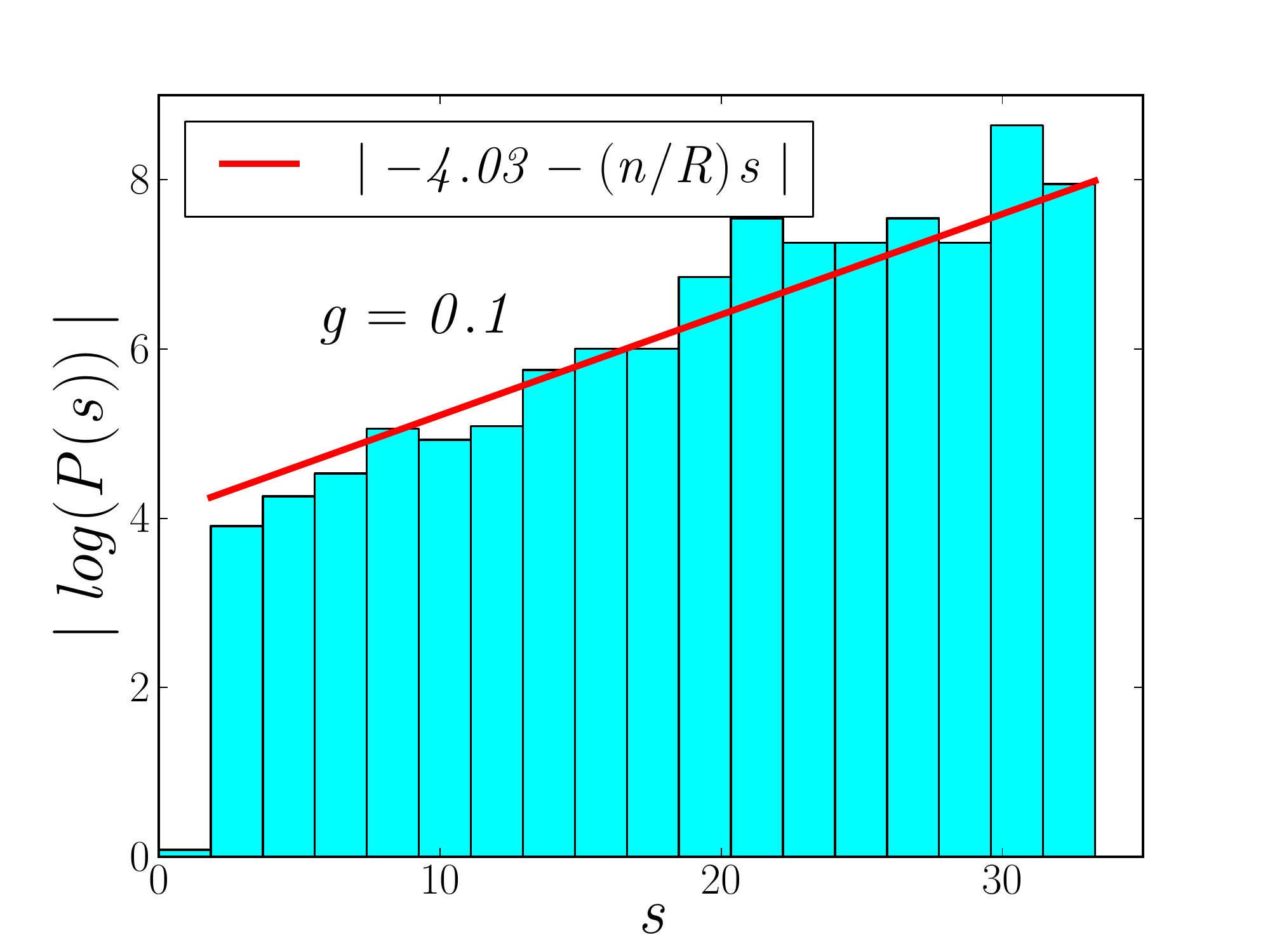}
\caption{\label{fig:P(s)2} Absolute value of $log$ of spacing distribution of the unfolded eigenvalues of $\rho_A$ is plotted with the adjacent level spacing $s$, for $N=16$ sites at half-filling $g=0.1$, and is linear fitted with slope $n/R$. This confirms prediction from random matrix theory. ( Refer Fig \ref{fig:P(s)1}. )}
\end{figure}

But we can prove that by noting that the matrix $M$ is proportional to the reduced density matrix for $g=1$, where the system is non-interacting. In that case, as shown by various authors \cite{ingo, arres}, Wick decomposition allows eigenvalues of $\rho_A$ to be written in terms of eigenvalues of the correlation matrix $C_{nm} = Tr (\rho_A c_n^{\dagger}c_m)$. The eigenvalues $\mu_l$ of the correlation matrix are related to eigevalues $\epsilon_l$ by the realtion :
\begin{equation}
\epsilon_l = log(\frac{1-\mu_l}{\mu_l}), \hspace{5pt} 0<\mu_l<1
\end{equation}
Following F. Arres et.al \cite{arres}, the elements of the correlation matrix are given by :
\begin{equation}
C_{nm} = \frac{1}{N}\sum_{k} e^{2\pi i k(n-m)/N}
\end{equation}
where the sum is over all occupied values of $k$. We note that, since $k$ and $-k$ are both allowed in the sum, $C_{nm}$ is real. Hence the correlation matrix has a real symmetric Toeplitz structure. Therefore, each block of $\rho_A$, which is proportional to the matrix $M$, must have the same number of degrees of freedom as a real symmetric Toeplitz matrix.

Motivated by random matrix theory and the fact that the system behaves `chaotic' for intermediate values of $g$, we claim that the adjacent level spacing distribution for intermediate values of $g$ will be the same as that of a random matrix ensemble of block diagonal matrices where each block is a real symmetric Toeplitz matrix whose elements are drawn from some distribution with mean $0$ and variance $1$. Ensemble of random real symmetric Toeplitz matrices have been previously investigated and the spacing distribution is found to be Poisson distribution, and not the usual GOE spacing distribution \cite{Toep}. This is because the number of degrees of freedom of $\tilde{N}\times\tilde{N}$ real symmetric Toeplitz matrix is $\tilde{N}$ and not $O({\tilde{N}}^2)$. In our case, since sizes of blocks vary, we have many ensembles of random real symmetric Toeplitz matrices of various sizes. So the spacing distribution will not exactly be the Poisson distribution. Since size of matrix is the only parameter in a random matrix ensemble, blocks of same size will have same distribution of eigenvalues.  Also eigenvalue distributions of blocks of different sizes will overlap. Hence, probability of degeneracy of the eigenvalues will be very high. Thus in the spacing distribution, there will be a pronounced peak at zero spacing. Apart from the peak, the effect of having blocks of various sizes will be to slow down the Poisson like decay. So the exponent in Poisson distribution will pick up a constant factor. The amplitude of the Poisson like decay function will be much smaller than the peak at zero spacing. Thus the spacing distribution we expect from random matrix theory is highly peaked at zero and having an exponentially decaying tail, the peak being much higher than amplitude of the exponential function. The exponential decay scale depends, as we will see below (see Fig.4), on the ratio of the number of distinct eigenvalues and the rank that we have investigated (see Fig.2) earlier.

To check the above result with exact numerical calculations, we need to unfold the eigenvalue spectrum to get a non-trivial distribution, viz. the average spacing of the unfolded energy level should be independent of the system size.  We  define the unfolded eigenvalues $\{\tilde{\epsilon_i}\}$, using the analytical result of the spectrum given in Eq.\ref{nu2},
\begin{equation}
\tilde{\epsilon}_i = I_0(2\sqrt{\mid \frac{S}{2}(ln \hspace{2pt} \epsilon_i+\frac{S}{2})\mid} ) \label{e_i}
\end{equation}
The adjacent level spacing distribution is then given by, 
\begin{equation}
P(s) = \sum_i \delta(s - \frac{|\tilde{\epsilon}_i - \tilde{\epsilon}_{i+1}|}{\Delta}).
\end{equation}
where $\Delta$ is the mean adjacent level spacing of $\tilde{\epsilon}_i$. Even though the distribution of $\epsilon_i$ is constrained by $\sum_i \epsilon_i =1$, the above distribution is unconstrained because $\tilde{\epsilon}_i$ diverges as $\epsilon_i \rightarrow 0$. The spacing distribution for $N=16$ sites is shown in Fig.\ref{fig:P(s)1} for various values of $g$. The distribution has a pronounced peak at zero spacing, but it goes down very fast to a peak of a single value, i.e, there is only one value corresponding to that spacing. After that, of course, there can be no decay. Thereafter, the peaks of single values occur at large intervals, the interval size also increasing with spacing, which means the effective probability of finding a spacing at a given interval keeps on decreasing. This discrete tail, of course, cannot be described by a continuous function. However, our understanding from random matrix gives the decay after the peak at zero spacing till the first peak of single value for intermediate values of $g$. The constant factor in the exponent is phenomenologically found to be equal to the fraction of distinct eigenvalues $n/R$, which would of course be equal to $1$ if there were no degeneracy. Fig. \ref{fig:P(s)2} gives the semilog plot. At $g=0$ and $g=1$, the spacing distributions are different because there the system is `integrable'. There the spacing distribution may still be Poisson-like with a different constant factor in the exponent, but no conclusion can be drawn at present. 

Away from the half-filling case, none of the above measures show much variation with the correlation factor $g$, which goes well with previous results from calculation of entanglement entropy \cite{Archak}.  

\section{Conclusion}
We have investigated eigenvalue spectrum of the reduced density matrix, as a function of the correlation factor $g$ for Gutzwiller state at half-filling. We have found that the rank $R$ and the largest eigenvalue $\epsilon_{max}$ of the reduced density matrix vary in such a way that their product clearly demarcates the insulating and metallic regions. In the insulating region it increases rapidly with $g$ and becomes almost constant in the metallic region. The fraction of distinct eigenvalues $n/R$ exhibits a `chaotic' behaviour  in the crossover region, and it shows a `integrable' behaviour at both insulating and metallic ends.

We have shown that the integrated entanglement spectrum deviates from the Calabrese-Lefevre result in the crossover region. We have related the deviation to existence of a finite correlation length and have shown that replacing the central charge by an effective finite-size corrected central charge, or, equivalently, writing the Calabrese-Lefevre result in terms of entanglement entropy gives the correct behaviour. Similar work has recently been done in the context of Heisenberg model \cite{XXZ}.
 
We have investigated the spacing distribution of non-zero eigenvalues of the reduced density matrix and have provided an understanding of it in terms of random matrices. The adjacent level spacing distribution of unfolded non-zero eigenvalues for intermediate values of Gutzwiller projection parameter $g$ is the same as that of an ensemble of random matrices obtained by replacing each block of reduced density matrix by a random real symmetric Toeplitz matrix. It is strongly peaked at zero, with an exponential tail proportional to $e^{-(n/R)s}$, where $s$ is the adjacent level spacing, the proportionality constant being much smaller than the height of the peak at zero spacing.

All our results relating to the crossover region are essentailly finite-size effects. This is because in the thermodynamic limit, for intermediate values of $g$, the ratio of the finite correlation length and the subsystem size goes to zero, and hence the system behaves exactly the non-interacting $g=1$ case for any non-zero value of $g$ \cite{metzner,Archak}. But as discussed in the introduction, finite-size effects remain significant at least as long as $N<10^5$, which are system sizes realizable in nanochains.

\end{document}